\documentclass[
floatfix,
aps,
prb,
amsmath,
twocolumn,
superscriptaddress, 
tightenlines,
]{revtex4-1}

\usepackage{graphicx}
\usepackage{dcolumn}
\usepackage{amsmath}
\usepackage{times}
\usepackage{amsfonts}
\usepackage{bm}
\usepackage{epsfig}
\newcommand{\pc}{{\cal P}_{\omega}}
\newcommand{\be}{\begin{equation}}
\newcommand{\ee}{\end{equation}}
\newcommand{\bea}{\begin{eqnarray}}
\newcommand{\eea}{\end{eqnarray}}

\def\ao{\alpha_{\omega}}
\def\bo{\beta_{\omega}}

\def\Eac{\mathcal{E}_{\mathrm{ac}}}

\def\xt{\mathcal{X}_{2}}
\def\eac{\epsilon}

\def\epseff{\varepsilon_{\mathrm{eff}}}
\def\pac{\delta}

\def\oc{\omega_{\mbox{\scriptsize {c}}}}

\def\tq{\tau_{\mbox{\scriptsize {q}}}}

\def\tem{\tau_{\mbox{\scriptsize {em}}}}

\newcommand{\req}[1]{Eq.\,(\ref{#1})}

\newcommand{\rfig}[1]{Fig.\,\ref{#1}}

\begin{document}

\title{
Microwave photoresistance in a 2D electron gas in separated Landau levels
}

\author{A.\,T. Hatke}
\affiliation{School of Physics and Astronomy, University of Minnesota, Minneapolis, Minnesota 55455, USA}

\author{M.\,A. Zudov}
\email[Corresponding author: ]{zudov@physics.umn.edu}
\affiliation{School of Physics and Astronomy, University of Minnesota, Minneapolis, Minnesota 55455, USA}

\author{L.\,N. Pfeiffer}
\affiliation{Princeton University, Department of Electrical Engineering, Princeton, NJ 08544, USA}

\author{K.\,W. West}
\affiliation{Princeton University, Department of Electrical Engineering, Princeton, NJ 08544, USA}


\begin{abstract}
Theories of microwave-induced resistance oscillations in high-mobility two-dimensional electron gas predict that with decreasing oscillation order $n$ {\em or} with increasing frequency $\omega$ the photoresistance maxima should appear closer to the cyclotron resonance harmonics due to increased Landau level separation.
In this experimental study we demonstrate that while for a given $\omega$ the peaks do move towards the harmonics with decreasing $n$, there is no corresponding movement with increasing $\omega$ for a given $n$.
These findings show that the positions of the photoresistance maxima cannot be directly linked to the Landau level separation challenging our current understanding of the phenomenon. 

\end{abstract}
\maketitle

Magnetotransport in high Landau levels of two-dimensional electron systems (2DESs) exhibits a variety of remarkable phenomena, such as microwave-(MIRO),\cite{miro:exp,zudov:2004,studenikin:2005,studenikin:2007,miro:th:disp,miro:th:inel} phonon-, \cite{piro:exp,piro:th} Hall field-\cite{hiro:exp,hiro:th} induced resistance oscillations, and several classes of combined oscillations.\cite{comb:exp,khodas:2010,comb:th}
Experimentally, all these effects often extend into the regime of separated Landau levels where even more phenomena, such as radiation-induced zero-resistance states,\cite{zrs:exp,zrs:th} dc field-induced zero-differential resistance states,\cite{zdrs:exp} and a sharp photoresistivity peak near the second harmonic of the cyclotron resonance,\cite{X2peak:exp} emerge. 
On the other hand, the majority of the theoretical proposals\cite{miro:th:inel,miro:th:disp,piro:th,hiro:th,comb:th} focus on the overlapping Landau-level regime and, as a result, their direct applicability to many experiments remains uncertain.

Theoretically, two mechanisms are usually discussed in relation to MIRO, {\em displacement}\cite{miro:th:disp,vavilov:2004} and {\em inelastic}.\cite{dmitriev:2005,miro:th:inel}
The displacement contribution\cite{miro:th:disp,vavilov:2004} originates from the modification of impurity scattering by microwave radiation, while the inelastic mechanism\cite{dmitriev:2005,miro:th:inel} owes to the microwave-induced nonequilibrium distribution of electrons.
In both cases, MIRO are understood in terms of optical transitions between the disorder-broadened Landau levels. 
In the regime linear in microwave intensity and overlapping Landau levels, the oscillatory photoresistivity can be described by\cite{dmitriev:2005}
\be
\frac {\delta \rho_\omega(\eac)}{\rho_0} \simeq - \eta {\pc} \lambda^2 \eac \sin 2\pi\eac\,.
\label{miro}
\ee
Here, $\rho_0$ is the resistivity at $B=0$, $\eac = \omega/\oc$, $\omega=2\pi f$ is the microwave frequency, $\oc=e B/m^*$ is the cyclotron frequency of an electron with an effective mass $m^*$, $\lambda = \exp(-\pi\ao\eac)$ is the Dingle factor, $\ao=(\omega\tq)^{-1}$, $\tq$ is the quantum lifetime, $\eta$ is a scattering parameter,\cite{n1} and $\pc$ is the dimensionless parameter proportional to the microwave power which, for circular polarization, is given by\cite{khodas:2010}
\be
\pc(\eac)= \frac {\pc^{0}} {(1 - \eac^{-1})^2+\bo^2},~\pc^{0}=\frac{e^2\Eac^2v_F^2}{\epseff \hbar^2 \omega^4}\,,
\label{pc2}
\ee
where $\bo\equiv(\omega\tem)^{-1}$, $\tem^{-1}=n_e e^2/2\sqrt{\epseff}\epsilon_0m^*c$, $2\sqrt{\epseff}=\sqrt{\varepsilon}+1$, $\varepsilon=12.8$ is the dielectric constant of GaAs, $v_F$ is the Fermi velocity, and $\Eac$ is the external (unscreened) microwave electric field. 

The photoresistance vanishes at the harmonics of the cyclotron resonance, $\eac=n=1,2,3,...$, and the positions of the MIRO maxima $(\eac^+)$ and minima $(\eac^-)$ are given by 
\be
\eac_n^{\pm} = n\mp\pac_n\,,
\label{eq.max}
\ee
where $\delta_n\equiv|\eac_n^{\pm}-n|$ is usually called the {\em phase}.
In a typical high-mobility 2DES, $\tq\sim\tem\sim 10^{-11}$ s and $\ao\sim\bo \ll 1$ at $f\sim 10^{11}$ Hz.
As a result, for all $n\neq1$, \req{miro} predicts $\pac_n \simeq  1/4$. 
However, close to the cyclotron resonance, the phase can become significantly smaller at higher $\omega$ due to strong enhancement of $\pc$ near the cyclotron resonance.

In the regime of separated Landau levels, \req{miro} is no longer valid and
the phase $\pac_n$ will be governed by the ratio of the Landau level width $\Gamma$ to the cyclotron energy, 
\be
\pac_n \simeq \frac {\kappa \Gamma}{\hbar\oc} \simeq \frac {\kappa \Gamma}{\hbar\omega}\cdot n\,.
\label{phase}
\ee
Here, $\kappa \sim 1$\cite{n2} and the last approximation in \req{phase} is justified at $\Gamma \ll \hbar\oc$.
To illustrate the origin of \req{phase} we consider, as an example, the leading part of the displacement contribution\cite{n3,dmitriev:2009a}
\be
\delta \rho_\omega(\eac) \propto \partial_{\omega} \langle \nu_{\varepsilon}\nu_{\varepsilon+\hbar\omega} \rangle_\varepsilon\,,
\label{disp}
\ee 
where $\nu_\varepsilon$ is the density of states at energy $\varepsilon$ and $\langle \dots \rangle_\varepsilon$ denotes averaging over the cyclotron energy, $\hbar\oc$.\cite{n4}
At $\Gamma < \hbar\oc$, the photoresitivity $\delta \rho_\omega(\eac)$ will be substantial only when the initial ($\nu_\varepsilon$) and the final ($\nu_{\varepsilon+\hbar\omega}$) densities of states overlap. 
As a result, the detuning from the closest cyclotron resonance harmonic must be close to $\Gamma$, $\hbar|\omega - n \oc| \sim \Gamma$, i. e. the condition equivalent to \req{phase}.

Equation \ref{phase} predicts that in the regime of separated Landau levels the phase $\pac_n$ should decrease when one lowers the oscillation order $n$ or raises the microwave frequency $\omega$. 
It also suggests that the evolution of the phase with the magnetic field should yield direct information on the $B$-dependence of $\Gamma$, which is {\em not} readily available from conventional transport measurements.
However, as we show below, our understanding of the phenomenon needs to be further improved before one attempts to extract $\Gamma$ from \req{phase}.

In this Rapid Communication we systematically examine the phase of MIRO over a wide range of microwave frequencies, covering both the overlapping and separated Landau level regimes.
We find that for a given frequency $\omega$ the phase of high order ($n\gtrsim 3$) MIRO is close to $1/4$, in agreement with \req{miro}, and is significantly smaller for lower orders, in agreement with \req{phase} and previous studies.\cite{zudov:2004,studenikin:2005,studenikin:2007}
However, we observe {\em no} decrease of $\pac_n$ with increasing $\omega$ within the accuracy of our measurements; for any given $n$, the phase remains constant over the whole range of frequencies studied. 
This finding contradicts \req{phase} indicating that the phase reduction commonly observed at low order MIRO\cite{zudov:2004,studenikin:2005,studenikin:2007} cannot be explained by existing theories of microwave photoconductivity.

\begin{figure}[t]
\includegraphics{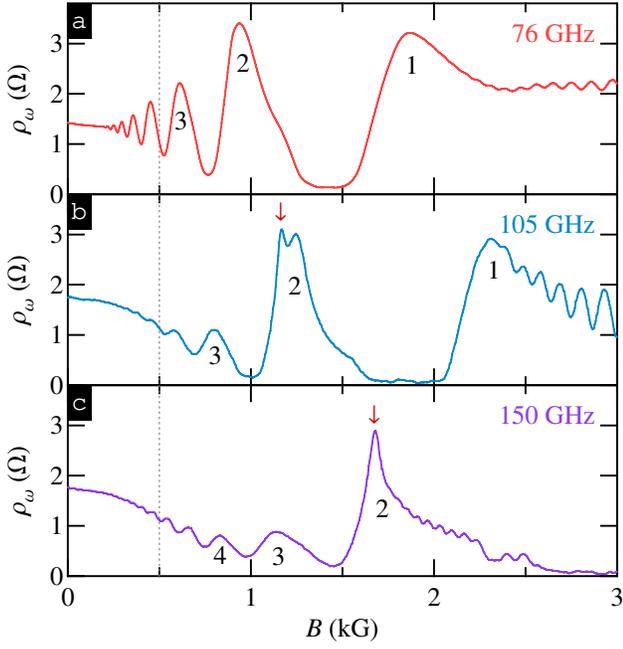}
\vspace{-0.1 in}
\caption{(Color online)
Magnetoresistivity $\rho_\omega(B)$ under microwave irradiation of (a) $f=76$ GHz, (b) $f=105$ GHz, and (c) $f=150$ GHz. 
Photoresistance peaks are marked by integers indicating the closest cyclotron resonance harmonic.
The vertical line marks the transition from overlapping to separated Landau levels estimated from $\oc\tq = \pi/2$.\cite{ando:1974b,laikhtman:1994}
}
\vspace{-0.2 in}
\label{fig1}
\end{figure}

Our sample is a Hall bar (width $w=100$ $\mu$m) cleaved from a GaAs/Al$_{0.24}$Ga$_{0.76}$As 300 \AA-wide quantum well grown by molecular beam epitaxy.
The density $n_e$ and the mobility $\mu$ were $3.6 \times 10^{11}$ cm$^{-2}$ and $\simeq 1.0 \times 10^7$ cm$^2$/Vs, respectively. 
Microwave radiation of frequency $f$ (60 GHz to 180 GHz), generated by Gunn and backward wave oscillators, was delivered to the sample via either a WR-28  waveguide or a 1/4-in.-diam light pipe. 
The microwave intensity was kept sufficiently low to ensure that all measurements were performed in the regime linear in microwave power.\citep{n6} 
The resistivity $\rho_\omega$ was measured at $T \simeq 0.5$ K under continuous microwave irradiation using a standard low-frequency lock-in technique.

In \rfig{fig1}\,(a)-1(c) we present magnetoresistivity $\rho_\omega(B)$ under microwave irradiation of (a) $f=76$ GHz, (b) $f=105$ GHz, and (c) $f=150$ GHz.
All three data sets exhibit pronounced MIRO extending over a progressively wider range of the magnetic fields with increasing microwave frequency, as prescribed by \req{miro}.
The data obtained at $f=105$ GHz also show that the photoresistivity near the second harmonic of the cyclotron resonance, $\omega/\oc=2$, clearly reveals a double-peak structure.
We attribute the sharper, lower $B$ feature [cf. $\downarrow$ in \rfig{fig1}\,(b)], to the recently reported so-called $\xt$ peak.\cite{X2peak:exp} 
As a result of its characteristic frequency dependence,\cite{X2peak:exp} this peak is not observed in our 2DES at lower frequencies [cf. \rfig{fig1}\,(a)] and becomes dominant at higher frequencies [cf. $\downarrow$ in \rfig{fig1}\,(c)].
In what follows we systematically investigate the positions of {\em all} the photoresistivity maxima, including the $\xt$ peak.

\begin{figure}[t]
\includegraphics{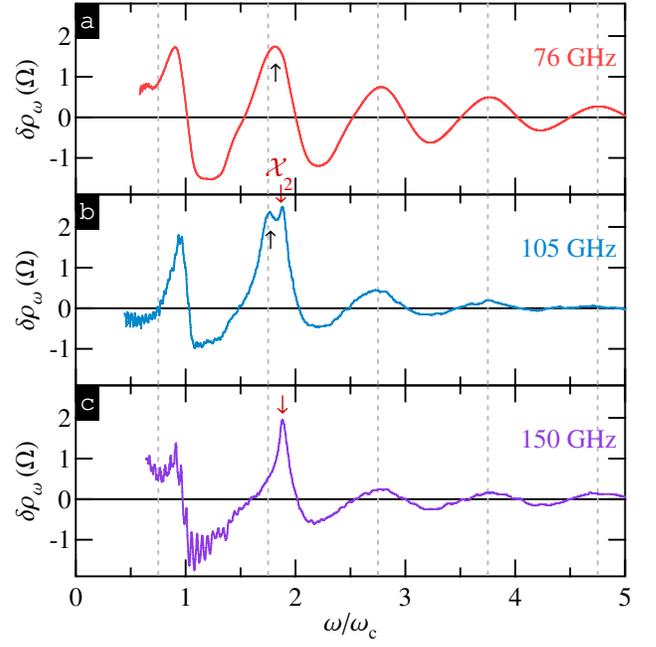}
\vspace{-0.1 in}
\caption{(Color online)
Microwave photoresistivity $\delta\rho_\omega $ versus $\omega/\oc$ for (a) $f=76$ GHz, (b) $f=105$ GHz, and (c) $f=150$ GHz.
The solid vertical lines correspond to $\omega/\oc=n-1/4$.
}
\vspace{-0.2 in}
\label{fig2}
\end{figure}
We start by extracting the oscillatory part of the resistivity $\delta\rho_\omega$ from the data in \rfig{fig1} and presenting the result as a function of $\omega/\oc \propto 1/B$ in \rfig{fig2}.
Plotted in such a way, the data readily reveal for {\em all} microwave frequencies that higher order ($n \ge 3$) MIRO peaks are well described by $\omega/\oc = n - \pac_n$, $\pac_n \simeq 1/4$ (cf. vertical lines), in agreement with \req{miro}.
This observation is in contrast to the lower order ($n =1,2$) peaks which exhibit considerably reduced phase values.
As discussed above, the phase reduction is anticipated in the regime of separated Landau levels, regardless of the physical mechanism or the shape of the Landau level.
More specifically, \req{phase} predicts that for a given (high enough) frequency $\omega$ the phase $\pac_n$ should decrease with decreasing $n$.
This result is consistent with our observations, as $\pac_1 <\pac_2 <\pac_3$.  
At the same time, \req{phase} also prescribes that for a given (low enough) cyclotron resonance harmonic $n$, the phase should monotonically {\em decrease} with increasing microwave frequency $\omega$, as $\pac_n \propto \Gamma/\omega$.
However, as we show next, our experimental findings fail to confirm this expectation.

\begin{figure}[t]
\includegraphics{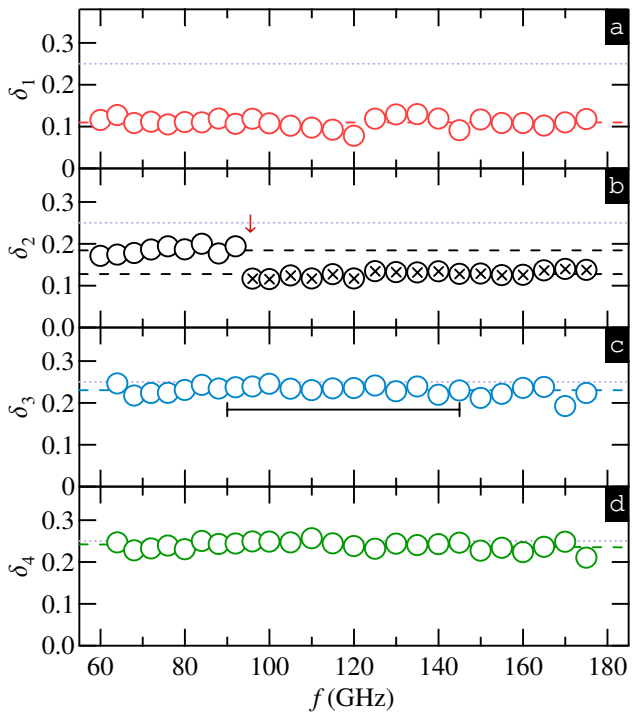}
\vspace{-0.1 in}
\caption{(Color online) Phase $\pac_n$ of the $\delta\rho_\omega$ maxima versus $f$ for (a) $n=1$, (b) $n=2$, (c) $n=3$, and (d) $n=4$.
The dashed lines represent average values, $\left < \pac_n \right >$: (a) $0.110$, (b) $0.184$ and $0.128$, (c) $0.230$, and (d) $0.243$.
The horizontal line in (c) is drawn at $\left < \pac_2\right > =0.184$ over the range of $B$ to the left-hand side of $\downarrow$ in (b).
}
\vspace{-0.2 in}
\label{fig3}
\end{figure}
Using the photoresistivity data such as that shown in \rfig{fig2} we extract the peak positions for all frequencies studied. 
The results are presented in \rfig{fig3}\,(a)-3(d) displaying the phase $\pac_n$ versus microwave frequency $f$ for (a) $n=1$, (b) $n=2$, (c) $n=3$, and (d) $n=4$.
For $n=3$ and $n=4$ [cf. \rfig{fig3}\,(c) and 3(d), respectively] we observe that the phase shows very little variation with microwave frequency and is close to 1/4 (cf. solid lines), a theoretical value expected in the regime of overlapping Landau levels-see \req{miro}.\cite{n5}
Indeed, the dashed lines drawn at average phase values, $\left < \pac_3\right > =0.230$ and $\left < \pac_4\right > =0.243$, show no sign of decrease at higher frequencies.
On the other hand, for the peak near $n=2$ [cf. \rfig{fig3}\,(b)] the phase clearly shows a jump occurring near 100 GHz in our 2DES.
This jump marks the appearance of the $\xt$ peak,\cite{X2peak:exp} which is absent at lower frequencies but dominates the response at higher frequencies.
Apart from this jump, the phase again shows little change with increasing frequency both for the MIRO peak ($f \leq 92$ GHz) and for the $\xt$ peak ($f \geq 96$ GHz). 

The average value of the phase at the $n=2$ MIRO peak is reduced considerably compared to $n\gtrsim 3$ MIRO peaks averaging at $\left < \pac_2 \right > = 0.184$ (cf. higher dashed line) and the phase of the $\xt$ peak is even lower averaging at $\simeq 0.128$ (cf. lower dashed line).  
Examination of the phase of the fundamental ($n=1$) MIRO peak in \rfig{fig3}\,(a) reveals a further reduced phase, $\left < \pac_1 \right > = 0.110$ (cf. dashed line) which again is almost {\em independent} of the microwave frequency.
Somewhat larger fluctuations of $\pac_1$ about the average value can be attributed to the overlap with the Shubnikov-de Haas oscillations and to the unavoidable variations in the incident microwave power which might affect the phase through the enhancement of $\pc$ near the cyclotron resonance-see \req{pc2}.

The ratio of the magnetic fields for the second and the third MIRO maxima can be estimated as $B_2/B_3 \simeq (3-\left < \pac_3 \right >)/(2-\left < \pac_2 \right >) \simeq 1.5$.
This value is about a factor of two lower than the variation in microwave frequency in our experiment and, therefore, the phases $\pac_2$ and $\pac_3$ can be directly compared over the same range of magnetic fields.
Indeed, the lower frequency range (below the onset of the $\xt$ peak), where the phase of the second MIRO peak can be reliably determined, can be mapped to a frequency range for the third MIRO peak.
This range is represented in \rfig{fig3}\,(c) by a horizontal line drawn between 90 GHz and 145 GHz.
It is clear that not only within this range but also at higher frequencies the phase of the third peak $\left < \pac_3 \right > =0.230$ remains higher than $\left < \pac_2 \right >$ = 0.184. 

\begin{figure}[t]
\includegraphics{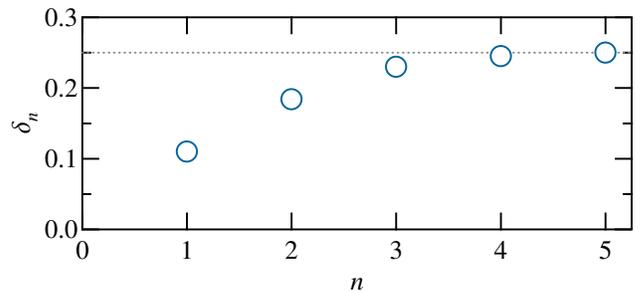}
\vspace{-0.1 in}
\caption{(Color online) Phase values obtained by averaging over all microwave frequencies vs the peak order $n$.
}
\vspace{-0.2 in}
\label{fig4}
\end{figure}

Taken together, our findings bring us to the conclusion that in our 2DES the phase of MIRO is determined primarily by the oscillation order $n$.
This result is summarized in \rfig{fig4}, showing average phase values as a function of $n$.
We notice that a very similar dependence was previously observed for $f = 57$ GHz.\cite{zudov:2004}
In the present Rapid Communication, we demonstrate that this dependence is universal, i.e., the phase values are not influenced by the microwave frequency.

In summary, we have studied the microwave-induced resistance oscillations and the novel $\xt$ peak in a high-mobility 2DES over a wide range of microwave frequencies.
For each microwave frequency, we have found that the phase of the lower-order MIRO becomes smaller with decreasing order $n$, consistent with earlier experiments.\cite{zudov:2004,studenikin:2005,studenikin:2007}
However, aside from an abrupt phase change near the second harmonic of the cyclotron resonance associated with the appearance of the $\xt$ peak,\cite{X2peak:exp} the phase of {\em all} photoresistance maxima, including the $\xt$ peak, is found to be {\em independent} of the microwave frequency and, thus, of the magnetic field.
These findings contradict the generally accepted view that in the regime of separated Landau levels the phase value directly reflects the ratio of the Landau level width to the cyclotron energy and therefore should decrease with the magnetic field.

We thank M. Dyakonov, M. Khodas and D. Polyakov for discussions and S. Hannas, G. Jones, J. Krzystek, T. Murphy, E. Palm, J. Park, D. Smirnov, and A. Ozarowski for technical assistance.
A portion of this work was performed at the National High Magnetic Field Laboratory, which is supported by NSF Cooperative Agreement No. DMR-0654118, by the State of Florida, and by the DOE.
The Minnesota group acknowledges support by the DOE Grant No. DE-SC002567 (high frequency measurements at NHMFL) and by the NSF Grant No. DMR-0548014 (low frequency measurements at Minnesota). 
The work at Princeton was partially funded by the Gordon and Betty Moore Foundation as well as the NSF MRSEC Program through the Princeton Center for Complex Materials (DMR-0819860).


\begin{thebibliography}{25}
\expandafter\ifx\csname natexlab\endcsname\relax\def\natexlab#1{#1}\fi
\expandafter\ifx\csname bibnamefont\endcsname\relax
 \def\bibnamefont#1{#1}\fi
\expandafter\ifx\csname bibfnamefont\endcsname\relax
 \def\bibfnamefont#1{#1}\fi
\expandafter\ifx\csname citenamefont\endcsname\relax
 \def\citenamefont#1{#1}\fi
\expandafter\ifx\csname url\endcsname\relax
 \def\url#1{\texttt{#1}}\fi
\expandafter\ifx\csname urlprefix\endcsname\relax\def\urlprefix{URL }\fi
\providecommand{\bibinfo}[2]{#2}
\providecommand{\eprint}[2][]{\url{#2}}

\bibitem[{mir({\natexlab{a}})}]{miro:exp}
\bibinfo{note}{
M. A. Zudov, R. R. Du, J. A. Simmons, and J. L. Reno, Phys. Rev. B {\bf 64}, 201311(R) (2001);
P. D. Ye, L. W. Engel, D. C. Tsui, J. A. Simmons, J. R. Wendt, G. A. Vawter, and J. L. Reno, Appl. Phys. Lett. {\bf 79}, 2193 (2001);
R. G. Mani, V. Narayanamurti, K. von Klitzing, J. H. Smet, W. B. Johnson, and V. Umansky, Phys. Rev. B {\bf 69}, 161306 (2004);
S. I. Dorozhkin, J. H. Smet, V. Umansky, and K. von Klitzing, {\em ibid} {\bf 71}, 201306(R) (2005);
C. L. Yang, R. R. Du, L. N. Pfeiffer, and K. W. West, {\em ibid.} {\bf 74}, 045315 (2006);
S. Wiedmann, G. M. Gusev, O. E. Raichev, T. E. Lamas, A. K. Bakarov, and J. C. Portal, {\em ibid.} {\bf 78}, 121301(R) (2008);
{\bf 80}, 035317 (2009);
{\bf 81}, 085311 (2010);
O. M. Fedorych, M. Potemski, S. A. Studenikin, J. A. Gupta, Z. R. Wasilewski, and I. A. Dmitriev, {\em ibid.} {\bf 81}, 201302 (2010);
I. V. Andreev, V. M. Murav'ev, I. V. Kukushkin, S. Schmult, and W. Dietsche, {\em ibid.} {\bf 83}, 121308(R) (2011);
A. T. Hatke, M. A. Zudov, L. N. Pfeiffer, and K. W. West, Phys. Rev. Lett. {\bf102}, 066804 (2009);
S. I. Dorozhkin, JETP Lett. {\bf 77}, 577 (2003); 
A. A. Bykov, {\em ibid.} {\bf 87}, 233 (2008);
{\bf 87}, 551 (2008);
{\bf 89}, 575 (2009);
{\bf 91}, 361 (2010);
A. A. Bykov and I. V. Marchishin, {\em ibid.} {\bf 92}, 71 (2010);
I. V. Andreev, V. M. Murav'ev, I. V. Kukushkin, J. H. Smet, K. von Klitzing, and V. Umanskii, {\em ibid.} {\bf 88}, 616 (2009).
}

\bibitem[{\citenamefont{Zudov}(2004)}]{zudov:2004}
\bibinfo{author}{\bibfnamefont{M.~A.} \bibnamefont{Zudov}},
 \bibinfo{journal}{Phys. Rev. B} \textbf{\bibinfo{volume}{69}},
 \bibinfo{pages}{041304(R)} (\bibinfo{year}{2004}).

\bibitem[{\citenamefont{Studenikin et~al.}(2005)\citenamefont{Studenikin,
 Potemski, Sachrajda, Hilke, Pfeiffer et~al.}}]{studenikin:2005}
\bibinfo{author}{\bibfnamefont{S.~A.} \bibnamefont{Studenikin}},
 \bibinfo{author}{\bibfnamefont{M.}~\bibnamefont{Potemski}},
 \bibinfo{author}{\bibfnamefont{A.}~\bibnamefont{Sachrajda}},
 \bibinfo{author}{\bibfnamefont{M.}~\bibnamefont{Hilke}},
 \bibinfo{author}{\bibfnamefont{L.~N.} \bibnamefont{Pfeiffer}},
 \bibnamefont{et~al.}, \bibinfo{journal}{Phys. Rev. B}
 \textbf{\bibinfo{volume}{71}}, \bibinfo{pages}{245313}
 (\bibinfo{year}{2005}).

\bibitem[{\citenamefont{Studenikin et~al.}(2007)\citenamefont{Studenikin,
 Sachrajda, Gupta, Wasilewski, Fedorych et~al.}}]{studenikin:2007}
\bibinfo{author}{\bibfnamefont{S.~A.} \bibnamefont{Studenikin}},
 \bibinfo{author}{\bibfnamefont{A.~S.} \bibnamefont{Sachrajda}},
 \bibinfo{author}{\bibfnamefont{J.~A.} \bibnamefont{Gupta}},
 \bibinfo{author}{\bibfnamefont{Z.~R.} \bibnamefont{Wasilewski}},
 \bibinfo{author}{\bibfnamefont{O.~M.} \bibnamefont{Fedorych}},
 \bibnamefont{et~al.}, \bibinfo{journal}{Phys. Rev. B}
 \textbf{\bibinfo{volume}{76}}, \bibinfo{pages}{165321}
 (\bibinfo{year}{2007}).

\bibitem[{mir({\natexlab{b}})}]{miro:th:disp}
\bibinfo{note}{V. I. Ryzhii, Sov. Phys. Solid State {\bf 11}, 2078 (1970);
V. I. Ryzhii, R. A. Suris, and B. S. Shchamkhalova, Sov. Phys. Semicond. {\bf 20}, 1299 (1986);
A. C. Durst, S. Sachdev, N. Read, and S. M. Girvin, Phys. Rev. Lett. {\bf 91}, 086803 (2003); 
J. Shi and X. C. Xie, {\em ibid.} {\bf 91}, 086801 (2003);
X. L. Lei and S. Y. Liu, {\em ibid.} {\bf 91}, 226805 (2003);
A. A. Koulakov and M. E. Raikh, Phys. Rev. B {\bf 68}, 115324 (2003);
V. Ryzhii and V. Vyurkov, {\em ibid.} {\bf 68}, 165406 (2003);
V. Ryzhii, {\em ibid.} {\bf 68}, 193402 (2003);
D.-H. Lee and J. M. Leinaas, {\em ibid.} {\bf 69}, 115336 (2004);
K. Park, {\em ibid.} {\bf 69}, 201301 (2004);
S. A. Mikhailov, {\em ibid.} {\bf 70}, 165311 (2004);
C. Joas, M. E. Raikh, and F. von Oppen, {\em ibid.} {\bf 70}, 235302 (2004);
C. Joas, J. Dietel, and F. von Oppen, {\em ibid.} {\bf 72}, 165323 (2005);
J. Dietel, L. I. Glazman, F. W. J. Hekking, and F. von Oppen, {\em ibid.} {\bf 71}, 045329 (2005);
M. Torres and A. Kunold, {\em ibid.} {\bf 71}, 115313 (2005);
J. Alicea, L. Balents, M. P. A. Fisher, A. Paramekanti, and L. Radzihovsky, {\em ibid.} {\bf 71}, 235322 (2005);
X. L. Lei and S. Y. Liu, {\em ibid.} {\bf 72}, 075345 (2005);
J. Dietel, {\em ibid.} {\bf 73}, 125350 (2006);
X. L. Lei, {\em ibid.} {\bf 73}, 235322 (2006);
{\em ibid.} {\bf 79}, 115308 (2009). 
O. E. Raichev, {\em ibid.} {\bf 81}, 165319 (2010);
V. Ryzhii, A. Chaplik, and R. Suris, JETP Lett. {\bf 80}, 363 (2004);
I. V. Pechenezhskii, S. I. Dorozhkin, and I. A. Dmitriev, {\em ibid.} {\bf 85}, 86 (2007); 
V. A. Volkov and E. E. Takhtamirov, JETP {\bf 104}, 602 (2007).
}

\bibitem[{\citenamefont{Vavilov and Aleiner}(2004)}]{vavilov:2004}
\bibinfo{author}{\bibfnamefont{M.~G.} \bibnamefont{Vavilov}} \bibnamefont{and}
  \bibinfo{author}{\bibfnamefont{I.~L.} \bibnamefont{Aleiner}},
  \bibinfo{journal}{Phys. Rev. B} \textbf{\bibinfo{volume}{69}},
  \bibinfo{pages}{035303} (\bibinfo{year}{2004}).

\bibitem[{mir({\natexlab{c}})}]{miro:th:inel}
\bibinfo{note}{I. A. Dmitriev, A. D. Mirlin, and D. G. Polyakov, Phys. Rev. Lett. {\bf 91}, 226802 (2003);
{\bf 99}, 206805 (2007);
Phys. Rev. B {\bf 70}, 165305 (2004);
{\bf 75}, 245320 (2007);
I. A. Dmitriev, M. Khodas, A. D. Mirlin, D. G. Polyakov, and M. G. Vavilov, {\em ibid.} {\bf 80}, 165327 (2009).
}

\bibitem[{pir({\natexlab{a}})}]{piro:exp}
\bibinfo{note}{
M. A. Zudov, I. V. Ponomarev, A. L. Efros, R. R. Du, J. A. Simmons, and J. L. Reno, Phys. Rev. Lett. {\bf 86}, 3614 (2001);
A. T. Hatke, M. A. Zudov, L. N. Pfeiffer, and K. W. West, {\em ibid.} {\bf 102}, 086808 (2009);
C. L. Yang {\em et al.}, Physica E (Amsterdam) {\bf 12}, 443 (2002);
A. A. Bykov, A. K. Kalagin, and A. K. Bakarov, JETP Lett. {\bf 81}, 523 (2005);
A. A. Bykov and A. V. Goran, {\em ibid} {\bf 90}, 578 (2009);
	A. T. Hatke, M. A. Zudov, L. N. Pfeiffer, and K. W. West, Phys. Rev. B, {\bf 84}, 121301 (2011).
}

\bibitem[{pir({\natexlab{b}})}]{piro:th}
\bibinfo{note}{I. V. Ponomarev and A. L. Efros, Phys. Rev. B {\bf 63}, 165305 (2001);
 X. L. Lei, {\em ibid.} {\bf 77}, 205309 (2008);
 O. E. Raichev, {\em ibid.} {\bf 80}, 075318 (2009).
 }

\bibitem[{hir({\natexlab{a}})}]{hiro:exp}
\bibinfo{note}{
C. L. Yang, J. Zhang, R. R. Du, J. A. Simmons, and J. L. Reno, Phys. Rev. Lett. {\bf 89}, 076801 (2002);
A. A. Bykov, J. Q. Zhang, S. Vitkalov, A. K. Kalagin and A. K. Bakarov, Phys. Rev. B {\bf 72}, 245307
 (2005);
W. Zhang, H.-S. Chiang, M. A. Zudov, L. N. Pfeiffer, and K. W. West, Phys. Rev. B {\bf 75}, 041304(R) (2007);
A. T. Hatke, M. A. Zudov, L. N. Pfeiffer, and K. W. West, {\em ibid.} {\bf 79}, 161308(R) (2009);
{\em ibid.} {\bf 83}, 081301(R) (2011);
Physicsa E (Amsterdam) {\bf 42}, 1081 (2010).
}

\bibitem[{hir({\natexlab{b}})}]{hiro:th}
\bibinfo{note}{
M. G. Vavilov, I. L. Aleiner, and L. I. Glazman, Phys. Rev. B {\bf 76}, 115331 (2007);
A. Auerbach and G. V. Pai, {\em ibid.} {\bf 76}, 205318 (2007);
X. L. Lei, Appl. Phys. Lett. {\bf 90}, 132119 (2007).
}

\bibitem[{com({\natexlab{a}})}]{comb:exp}
\bibinfo{note}{
W. Zhang, M. A. Zudov, L. N. Pfeiffer, and K. W. West, Phys. Rev. Lett. {\bf 98}, 106804 (2007); 
{\bf 100}, 036805 (2008);
Physica E (Amsterdam) {\bf 40}, 982 (2008);
A. T. Hatke, H.-S. Chiang, M. A. Zudov, L. N. Pfeiffer, and K. W. West, Phys. Rev. Lett. {\bf 101}, 246811 (2008);
Phys. Rev. B {\bf 77}, 201304(R) (2008);
M. A. Zudov, H.-S. Chiang, A. T. Hatke, W. Zhang, L. N. Pfeiffer, and K. W. West, Int. J. of Mod. Phys. B {\bf 23}, 2684 (2009).
}

\bibitem[{kho()}]{khodas:2010}
\bibinfo{note}{
M. Khodas, H. -S. Chiang, A. T. Hatke, M. A. Zudov, M. G. Vavilov, L. N. Pfeiffer, and K. W. West, {\em ibid.} {\bf 104}, 206801 (2010).
}

\bibitem[{com({\natexlab{b}})}]{comb:th}
\bibinfo{note}{M. Khodas and M. G. Vavilov, Phys. Rev. B {\bf 78}, 245319 (2008);
 X. L. Lei, {\em ibid.} {\bf 79}, 115308 (2009);
 I. A. Dmitriev, R. Gellmann, and M. G. Vavilov, {\em ibid.} {\bf 82}, 201311 (2010); 
 X. L. Lei, Appl. Phys. Lett. {\bf 91}, 112104 (2007);
 X. L. Lei and S. Y. Liu, {\em ibid.} {\bf 93}, 082101 (2008).
}

\bibitem[{zrs({\natexlab{a}})}]{zrs:exp}
\bibinfo{note}{
R. G. Mani, J. H. Smet, K. von Klitzing, V. Narayanamurti, W. B. Johnson, and V. Umansky, Nature (London) {\bf 420}, 646 (2002);
M. A. Zudov, R. R. Du, L. N. Pfeiffer and K. W. West, Phys. Rev. Lett. {\bf 90}, 046807 (2003);
{\bf 96}, 236804 (2006);
Phys. Rev. B {\bf 73}, 041303(R) (2006);
R. R. Du, M. A. Zudov, C. L. Yang, Z. Q. Yuan, L. N. Pfeiffer, and K. W. West, Int. J. of Mod. Phys. B {\bf 18}, 3465 (2004);
C. L. Yang, M. A. Zudov, T. A. Knuuttila, R. R. Du, L. N. Pfeiffer, and K. W. West, Phys. Rev. Lett. {\bf 91}, 096803 (2003);
R. G. Mani, J. H. Smet, K. von Klitzing, V. Narayanamurti, W. B. Johnson, and V. Umansky, {\em ibid.} {\bf 92}, 146801 (2004);
R. L. Willett, L. N. Pfeiffer and K. W. West, {\em ibid.} {\bf 93}, 026804 (2004);
J. H. Smet, B. Gorshunov, C. Jiang, L. Pfeiffer, K. West, V. Umansky, M. Dressel, R. Meisels, F. Kuchar, and K. von Klitzing {\em ibid.} {\bf 95}, 116804 (2005);
R. R. Du, M. A. Zudov, C. L. Yang, L. N. Pfeiffer, and K. W. West, Physica E (Amsterdam) {\bf 22}, 7 (2004);
A. A. Bykov, I. V. Marchishin, A. V. Goran, and D. V. Dmitriev, Appl. Phys. Lett. {\bf 97}, 082107 (2010);
S. I. Dorozhkin, L. N. Pfeiffer, K. W. West, K. von Klitzing, and J. H. Smet, Nature Phys. {\bf 7}, 336 (2011).
}

\bibitem[{zrs({\natexlab{b}})}]{zrs:th}
\bibinfo{note}{
A. V. Andreev, I. L. Aleiner and A. J. Millis, Phys. Rev. Lett. {\bf 91}, 056803 (2003);
A. Auerbach, I. Finkler, B. I. Halperin and A. Yacoby, {\em ibid.} {\bf 94}, 196801 (2005);
P. W. Anderson and W. F. Brinkman, arXiv:cond-mat/0302129;
I. G. Finkler and B. I. Halperin, Phys. Rev. B {\bf 79}, 085315 (2009).
}

\bibitem[{zdr()}]{zdrs:exp}
\bibinfo{note}{
A. A. Bykov, J.-Q. Zhang, S. Vitkalov, A. K. Kalagin, and A. K. Bakarov, Phys. Rev. Lett. {\bf 99}, 116801 (2007);
A. T. Hatke, H.-S. Chiang, M. A. Zudov, L. N. Pfeiffer, and K. W. West, Phys. Rev. B {\bf 82}, 041304(R) (2010).
}

\bibitem[{X2p()}]{X2peak:exp}
\bibinfo{note}{
Y. Dai, R. R. Du, L. N. Pfeiffer, and K. W. West, Phys. Rev. Lett. {\bf 105}, 246802 (2010);
A. T. Hatke, M. A. Zudov, L. N. Pfeiffer, and K. W. West, Phys. Rev. B {\bf 83}, 121301(R) (2011);
{\bf 83} 201301(R) (2011).
}

\bibitem[{\citenamefont{Dmitriev et~al.}(2005)\citenamefont{Dmitriev, Vavilov,
 Aleiner, Mirlin, and Polyakov}}]{dmitriev:2005}
\bibinfo{author}{\bibfnamefont{I.~A.} \bibnamefont{Dmitriev}},
 \bibinfo{author}{\bibfnamefont{M.~G.} \bibnamefont{Vavilov}},
 \bibinfo{author}{\bibfnamefont{I.~L.} \bibnamefont{Aleiner}},
 \bibinfo{author}{\bibfnamefont{A.~D.} \bibnamefont{Mirlin}},
 \bibnamefont{and} \bibinfo{author}{\bibfnamefont{D.~G.}
 \bibnamefont{Polyakov}}, \bibinfo{journal}{Phys. Rev. B}
 \textbf{\bibinfo{volume}{71}}, \bibinfo{pages}{115316}
 (\bibinfo{year}{2005}).

\bibitem[{n1()}]{n1}
\bibinfo{note}{In general, scattering parameter $\eta$ contains both the displacement and inelastic parts whose relative contributions depend on temperature and correlation properties of the disorder potential.}

\bibitem[{n2()}]{n2}
\bibinfo{note}{The value of $\kappa$ depends both on the underlying microscopic mechanism (e.g., displacement or inelastic) and on the Landau level shape (e.g., Lorentzian or Gaussian).}

\bibitem[{n3()}]{n3}
\bibinfo{note}{Similar arguments apply to the inelastic contribution.}

\bibitem[{\citenamefont{Dmitriev et~al.}(2009)\citenamefont{Dmitriev,
 Dorozhkin, and Mirlin}}]{dmitriev:2009a}
\bibinfo{author}{\bibfnamefont{I.~A.} \bibnamefont{Dmitriev}},
 \bibinfo{author}{\bibfnamefont{S.~I.} \bibnamefont{Dorozhkin}},
 \bibnamefont{and} \bibinfo{author}{\bibfnamefont{A.~D.}
 \bibnamefont{Mirlin}}, \bibinfo{journal}{Phys. Rev. B}
 \textbf{\bibinfo{volume}{80}}, \bibinfo{eid}{125418} (\bibinfo{year}{2009}).

\bibitem[{n4()}]{n4}
\bibinfo{note}{In overlapping Landau levels, $\nu_\varepsilon \propto 1 - 2\lambda \cos 2\pi\varepsilon$, $\partial_{\eac}\langle \nu_{\varepsilon}\nu_{\varepsilon+\eac} \rangle_\varepsilon = - 2 \lambda^2\sin 2\pi \eac$, and one obtains \req{miro}.}

\bibitem[{\citenamefont{Ando}(1974)}]{ando:1974b}
\bibinfo{author}{\bibfnamefont{T.}~\bibnamefont{Ando}}, \bibinfo{journal}{J.
 Phys. Soc. Jap.} \textbf{\bibinfo{volume}{37}}, \bibinfo{pages}{1233}
 (\bibinfo{year}{1974}).

\bibitem[{\citenamefont{Laikhtman and Altshuler}(1994)}]{laikhtman:1994}
\bibinfo{author}{\bibfnamefont{B.}~\bibnamefont{Laikhtman}} \bibnamefont{and}
 \bibinfo{author}{\bibfnamefont{E.~L.} \bibnamefont{Altshuler}},
 \bibinfo{journal}{Ann. Phys. (N.Y.)} \textbf{\bibinfo{volume}{232}},
 \bibinfo{pages}{332} (\bibinfo{year}{1994}).
 
\bibitem[{n6()}]{n6}
\bibinfo{note}{At elevated intensities, the phase is expected to decrease as
  $1/\sqrt{\pc}$.\cite{vavilov:2004,dmitriev:2005,n7}}

\bibitem[{n5()}]{n5}
\bibinfo{note}{Observation of a 1/4 phase at $n \gtrsim 3$ confirms that for all frequencies studied our measurements are performed in the regime linear in microwave intensity.}
 
\bibitem[{n7()}]{n7}
\bibinfo{note}{At high enough microwave power, the MIRO phase can be reduced due to, e.g. multiphoton processes, see e.g. A. T. Hatke, M. A. Zudov, M. Khodas, L. N. Pfeiffer, and K. W. West, Phys. Rev. B {\bf 84}, 241302(R) (2011).}

\end{thebibliography}
\end{document}